\pdfoutput=1
\documentclass{article}

\usepackage{iclr2027/iclr2027_conference}
\iclrfinalcopy

\usepackage[utf8]{inputenc} 
\usepackage[T1]{fontenc}    
\usepackage{hyperref}       
\usepackage{url}            
\usepackage{booktabs}       
\usepackage{amsfonts}       
\usepackage{nicefrac}       
\usepackage{microtype}      
\usepackage{xcolor}         

\usepackage{graphicx}
\usepackage{amsmath}
\usepackage{multicol}
\usepackage{wrapfig,sidecap}
\usepackage{algorithmic}
\usepackage{algorithm, setspace}

\usepackage{caption}

\title{Attention when you need}

\author{%
  Grayson~Matthew$^{1}$\thanks{Equal contribution.} \quad
  Lokesh~Boominathan$^{1}$\footnotemark[1] \quad
  Yizhou~Chen$^{1}$ \\
  \bfseries Matthew~McGinley$^{3,4,5}$ \quad
  Xaq~Pitkow$^{1,2}$ \\[4pt]
  \normalfont
  $^{1}$Neuroscience Institute, Carnegie Mellon University \\
  $^{2}$Department of Machine Learning, Carnegie Mellon University \\
  $^{3}$Department of Neuroscience, Baylor College of Medicine \\
  $^{4}$Jan and Dan Duncan Neurological Research Institute, Texas Children's Hospital \\
  $^{5}$Department of Electrical and Computer Engineering, Rice University \\[4pt]
  \normalfont\texttt{rcmatthe@andrew.cmu.edu, boominathanlokesh@gmail.com,} \\
  \normalfont\texttt{yizhoucc@gmail.com, mmcginle@bcm.edu, xaq@cmu.edu}
}

\begin{document}

\maketitle

\begin{abstract}
  Paying attention improves performance, but attention is metabolically costly, so how should a resource-efficient agent allocate it? We study optimal allocation strategies using a normative model of a signal detection task in which attention comes at a cost. The model reveals that optimal attention is temporally structured in one of two patterns depending on task conditions: when attention costs penalize intense focus, optimal attention ramps up as evidence for the signal accumulates, but with less prohibitive attention costs, optimal attention fluctuates rhythmically. In cases with rhythmic attention, allocation frequency increases with task parameters such as reward magnitude for successful detection, signal brevity, and signal frequency. We argue that rhythmic attention emerges naturally in an agent whose belief updates are dominated by a stable temporal prior and not by the likelihood of new sensory observations. Our results characterize the conditions under which rhythmic vs. ramping attention is optimal and offer a normative account of attentional fluctuations observed in sustained attention tasks.
\end{abstract}
 
\section{Introduction}
 
Good investment pays off. Across tasks, performance improves as the resources devoted to them, such as time, effort, money, and compute, increase from zero. These resources are not free, which raises the question of how an agent should allocate them to maximize performance while minimizing expenditure. In machine learning, this question motivates a large literature on conditional computation, which spends compute selectively rather than uniformly \citep{Scardapane_2024}: mixture-of-experts layers route each token to a subset of parameters \citep{shazeer2017outrageously}, early-exit and mixture-of-depths mechanisms vary how many layers an input passes through \citep{teerapittayanon2016branchynet, raposo2024mixture}, and adaptive computation time lets a network decide how long to process each input \citep{graves2016adaptive, banino2021pondernet}. The question becomes harder for agents that process streams, where allocation must be decided moment to moment without knowing what is coming: a video model deciding which frames to encode \citep{wu2019adaframe}, a recurrent network learning when to skip state updates \citep{campos2018skip}, or a reinforcement learning agent that must pay to observe its environment \citep{nam2021reinforcement, krale2023act}.
 
We study this allocation problem for a resource that brains and machines alike must budget: attention. The term has numerous meanings across machine learning and neuroscience, from self-attention in transformers to top-down and bottom-up attention in the brain. We use it in the sense closest to hard attention in machine learning \citep{mnih2014recurrent} and to attention as studied in neuroscience: a controllable resource that an agent invests to improve the quality of its observations. More attention improves performance when directed at the relevant features of a task \citep{cohen2010neuronal, cohen2011attention, de2017dynamic}, but this benefit comes at a cost. In the brain, tasks requiring sustained attention over long periods cause fatigue, leading to variable performance and an overall decline as time on task increases \citep{gilden1995nature, warm2008vigilance, fortenbaugh2017recent, mackworth1948breakdown, broadbent1971decision}. In artificial systems, the cost is paid in compute, latency, or energy.
 
We formalize this trade-off with a signal detection task, in which an agent must identify an ambiguous signal amid noise and chooses, at each moment, how much to pay for a more reliable observation. We designed the task to model sustained attention in animals, with timing chosen to match an existing rodent vigilance experiment \citep{de2022strategic}. However, the core task demands generalize naturally to any agent that must identify sparse, partially observed signals under resource constraints, from a driver waiting for a red light to turn green, to a trader watching the market for signs to buy or sell, to an artificial agent choosing which frames or tokens warrant more expensive, more capable computation.
 
To find solutions to our task, we develop a normative model, which reveals that optimal attention is never constant. Instead, it follows one of two consistent within-trial patterns, depending on task parameters. In the ``rhythmic'' pattern, attention is non-zero only sparsely, at regular intervals. In the ``ramping'' pattern, attention rises as evidence for the signal accumulates and drops when there is a lack of evidence. Which pattern emerges depends on how steeply attention costs grow with its informational benefit.

This result bears on a common design choice. Fixed-stride schedules, such as encoding every $k$th frame of a video, are widespread heuristics in systems that process streams. Our model identifies when such a schedule is not merely convenient but optimal, when it should give way to evidence-driven allocation, and how the optimal stride should scale with the task.

Our contributions are as follows:
\begin{itemize}
    \item We formulate sustained attention as a POMDP in which the agent pays to control the reliability of its observations, and we solve for optimal policies by value iteration over beliefs.
    \item We show that optimal attention allocation policies exhibit discrete jumps as a function of belief, producing two regimes, rhythmic and ramping, selected by the curvature of the attention cost.
    \item We show that rhythm emerges when belief updates between checks are dominated by the temporal prior rather than by new observations.
    \item We characterize how the rhythmic stride scales with reward, signal duration, and noise duration, and we show that optimal policies substantially outperform constant attention at matched mean attention.
\end{itemize}

\section{Related Work}
 
\textbf{Sustained attention.}
There are several classes of attention studied in neuroscience \citep{ocasio2011attention, lindsay2020attention}. We focus on sustained attention, also known as vigilance: a subject's preparedness to detect infrequent and unpredictable signals over extended periods \citep{sarter2001cognitive}. Despite the term ``sustained,'' there is evidence that attention fluctuates between optimal and suboptimal states over time, due to factors such as arousal, mind wandering, and competing behavioral demands \citep{esterman2014intrinsic, esterman2019models, busch2010spontaneous}. These fluctuations are often interpreted as lapses.
By assigning attention a cost, our work explores a setting in which attention that is suboptimal for \textit{performance} may optimally solve the \textit{trade-off} between the performance gained by paying attention and the price of paying it.
 
\textbf{Attention as a costly action.}
\citet{balkenius1999attention} argued that attention should be viewed as an action rather than solely a sensory process. Several works have adopted this perspective, building optimal control models that treat attention as an action choice to determine where \citep{bandera1996residual, paletta2005q, minut2001reinforcement} and how often \citep{di2012attentional} to attend. The same perspective underlies hard attention in deep learning, where a recurrent model learns through reinforcement where to direct a limited-resolution glimpse \citep{mnih2014recurrent}. Normative accounts in economics and cognitive science likewise model attention as a resource with explicit costs, deriving behavior from the trade-off between the value of information and the price of acquiring it \citep{howard1966information, sims2003implications, lieder2020resource}. In models of perceptual decision-making, costly evidence accumulation determines when an agent should stop sampling and decide \citep{wald1945sequential, frazier2007sequential, drugowitsch2012cost}; there, the quality of each sample is fixed, whereas our agent also chooses how much to invest in each one. The work most relevant to ours is \citet{chebolu2022vigilance}, which trained reinforcement learning agents on an abstract sustained attention task to study the trade-off between the benefits of high-quality observations and the attentional costs involved. 
We construct a conceptually similar model, but we extend the analysis to characterize patterns of attention and why they emerge.

\textbf{Costly observation in reinforcement learning.}
A line of work in reinforcement learning considers agents that must pay to observe their environment. In action-contingent noiselessly observable MDPs, the agent chooses at each step whether to pay for an exact observation of the state \citep{nam2021reinforcement}, and related work develops algorithms for this and closely related settings \citep{bellinger2021active, krale2023act}. These works aim to learn efficient policies; our aim is to characterize what optimal policies look like and why. Our formulation also differs in that observation quality is graded rather than all-or-nothing: the agent chooses how much to invest in each observation, not only whether to observe. Graded, noisy observations are what make the ramping regime possible. When a single observation resolves all uncertainty, there is no evidence to accumulate.
 
\textbf{Allocating computation over streams.}
Systems that process streams must decide when to spend computation, and two families of strategies recur. The first is fixed-stride sampling; video recognition models, for example, commonly process frames sampled at uniform intervals \citep{wang2016temporal, feichtenhofer2019slowfast}. The second adapts allocation to the input. Adaptive frame selection policies decide which frames merit processing \citep{wu2019adaframe, korbar2019scsampler}, recurrent networks can learn to skip state updates under a computational budget \citep{campos2018skip}, and cascades escalate from cheap to expensive models only when earlier stages find the input promising or uncertain, from the classic detection cascade of \citet{viola2001rapid} to cascades of large language models \citep{chen2023frugalgpt}. The same dichotomy appears in control theory as periodic versus event-triggered sampling, where event-triggered schemes can achieve lower estimation error at the same average sampling rate \citep{astrom2002comparison}. Our two regimes map onto these families. Rhythmic policies behave as fixed-stride schedules, and ramping policies escalate investment in stages as evidence accumulates, much like a cascade. In machine learning, the choice between these strategies is usually made empirically. Our model instead derives both as optimal solutions to a single problem and identifies the conditions that determine which is optimal: chiefly the curvature of the attention cost, and whether the agent's beliefs between observations are driven by its prior or by what it observes.
 
\section{Task and Model}
 
\subsection{Intuition} 

We consider an agent in a world whose state is latent: the agent cannot see the state directly and must infer it from observations. It therefore maintains a \textit{belief}, a posterior probability distribution over world states, which it updates as it \textit{observes} the world. In signal detection, the relevant distinction is binary: the world is in either a \textit{signal} or a \textit{noise} (non-signal) state, and the agent must report when a signal is present. To define an optimization objective, we assign values to the possible outcomes. Correctly detected signals (``hits'') contribute positively to reward, representing value gained by success at the task. Undetected signals (``misses'') forgo this value, and detections reported while the world is in a noise state (``false alarms'') incur a penalty, representing value lost by failure at the task.
 
\subsection{Implementation}
\begin{figure}[t]
    \centering
    \includegraphics[width=1\linewidth]{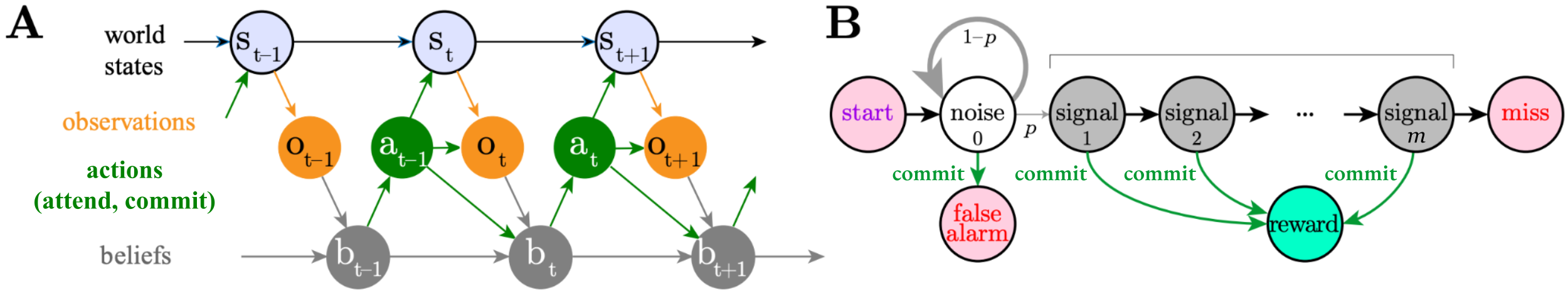}
    \caption{Formal task structure. ({\bf A}) In this POMDP, the agent constructs beliefs $b_t$ that reflect inferences about the latent world state $s_t$ using observations $o_t$ controlled by the ``attend'' action $a_t$. The agent uses these beliefs to plan ``commit'' actions that may obtain rewards. ({\bf B}) Specific world states and transitions in the finite state machine modeling this task. The noise state has a geometric duration. The signal lasts throughout a sequence of $m$ signal states, with a tighter distribution the higher $m$ is. Commit actions produce rewards when taken during the signal period and otherwise lead to a false alarm and penalty.}
    \label{POMDP}
\end{figure}

We model the task as a partially observable Markov decision process (POMDP; Fig~\ref{POMDP}A) \citep{kaelbling1998planning} in discrete time. At each step, the agent decides how much to attend and whether or not to report, or ``\emph{commit},'' to a signal's presence. 
 
We construct task trials as a sequence beginning in a start state. The start state transitions with 100\% probability to the noise state in the next step, and at each step thereafter, the noise state has a small, fixed probability $q$ of transitioning to the first signal state, so noise durations are geometrically distributed. The number of signal states $m$ can be varied to change the distribution of signal durations. Each signal state transitions to the next (or, for the final signal state, out of the signal period) with probability $p$, so signal durations follow a negative binomial distribution with mean $\frac{m}{p}$. For $m=1$, this distribution is geometric, like the distribution of noise durations; as $m$ increases with the mean held fixed, it concentrates around the mean. Task trials end in one of three ways. If the agent does not commit before the world transitions out of the final signal state, the trial ends as a miss. If the agent commits while the world is in the noise state, the trial is a false alarm. If the agent commits during any signal state, the trial is a hit (Fig. \ref{POMDP}B).
 
The agent maintains a belief $b_t$, a probability vector over world states, which it updates exactly at each step by Bayesian filtering (the forward recursion of a hidden Markov model), combining its previous belief, the known transition dynamics, and the likelihood of its current observation.
Attention increases the reliability of the agent's next observation, with higher attention yielding higher reliability but also higher cost. Specifically, observations are binary, either signal or noise, and for attention $a\ge0$, the observation matches the true state (signal or noise) with probability $\sigma(a)=\frac{1}{1+e^{-a}}$. At $a=0$, observations are correct with probability $\frac{1}{2}$ and therefore carry no information. The cost of attention is $C(a)=c_1a^{c_2}$, with scale $c_1>0$ and exponent $c_2>0$.
 
The agent maximizes expected discounted reward, combining attention costs, the hit reward, and the false alarm penalty. Because beliefs are sufficient statistics for the agent's history, the POMDP can be recast as a fully observed MDP over beliefs \citep{kaelbling1998planning}, which we solve by value iteration on a discretized belief space (Appendix~\ref{vimethods}). The resulting policy maps each belief to an attention level and a commit rule.
 
\section{Results}\label{results}
 
\subsection{Rhythmic and ramping attention emerge in optimal policies.}
 
\begin{figure}[t]
    \centering
    \includegraphics[width=\linewidth]{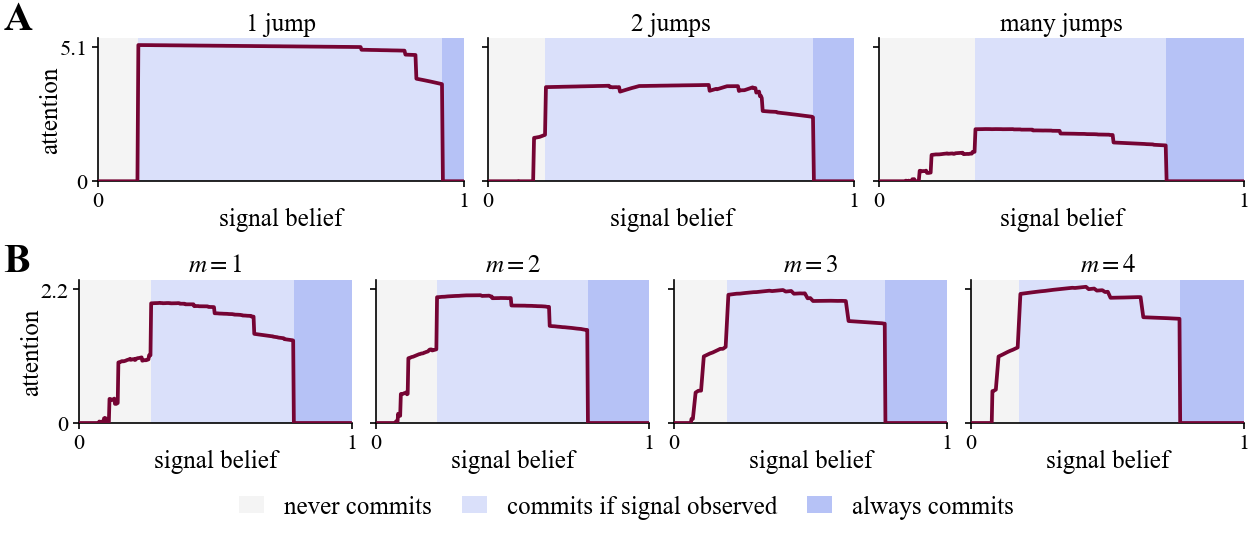}
    \caption{Attention policies over lower-dimension belief spaces. Attention over signal belief is shown in maroon; the background color shows if this signal belief is a level where the agent will not commit (light grey), will commit only if it observes signal (light blue), or will always commit (darker blue). ({\bf A}) When we model a geometric distribution for signal duration using a single signal state ($m=1$), belief is one-dimensional, so we may visualize full policies as a level of attention over total signal belief. We show a single-jump (left), 2-jump (middle) and many-jump (right) example here. ({\bf B}) We can visualize a \emph{slice} of policies with $m>1$ by fixing all but one signal state at 0. Here, we show an example of many-jumps policies where signal belief is concentrated in the final signal state for, from left to right, 1, 2, 3, and 4 signal states. We show and discuss more examples in the Appendix \ref{morevis}.}
    \label{fig:policy1}
\end{figure}
 
With one signal state ($m=1$), the belief space is one-dimensional, so we can visualize optimal policies in full (Fig. \ref{fig:policy1}A). These policies exhibit \textit{jumps} in optimal attention across signal belief. At low beliefs, no attention is paid because the signal is unlikely to be present; at intermediate beliefs, attention is allocated; and at high beliefs, attention again drops to zero because the signal is nearly certain, so attending to collect more information is unnecessary. Although policies become harder to visualize as $m$ grows, these jumps persist for higher numbers of signal states; we visualize up to $m=4$ in Fig. \ref{fig:policy1}B.
 
Together with the commit rule, these jumps partition belief space into regions of qualitatively different behavior. Where signal belief is low, the chance that the signal is present is too low to justify engaging, so the agent neither attends nor commits: it \emph{waits}. Where signal belief is high, attention is an added cost that would likely only confirm what the agent already believes, so it commits without attending. In the regions where the agent does attend, its commit rule takes one of two forms. Either it does not commit regardless of what it observes, \emph{gathering information} that moves its belief without ending the trial, or it commits only if it observes signal, using the observation to \emph{decide}. Transitions between these behaviors occur at jumps in the policy. If there is a single jump, the policy switches directly from waiting to deciding. If there are several, the policy switches from waiting to information gathering at the first jump, then from information gathering to deciding at the jump to maximum attention. The number of jumps is closely tied to the two patterns of attention allocation we observe across trials: rhythmic and ramping.

\begin{figure}[h]
    \centering
    \includegraphics[width=1\linewidth]{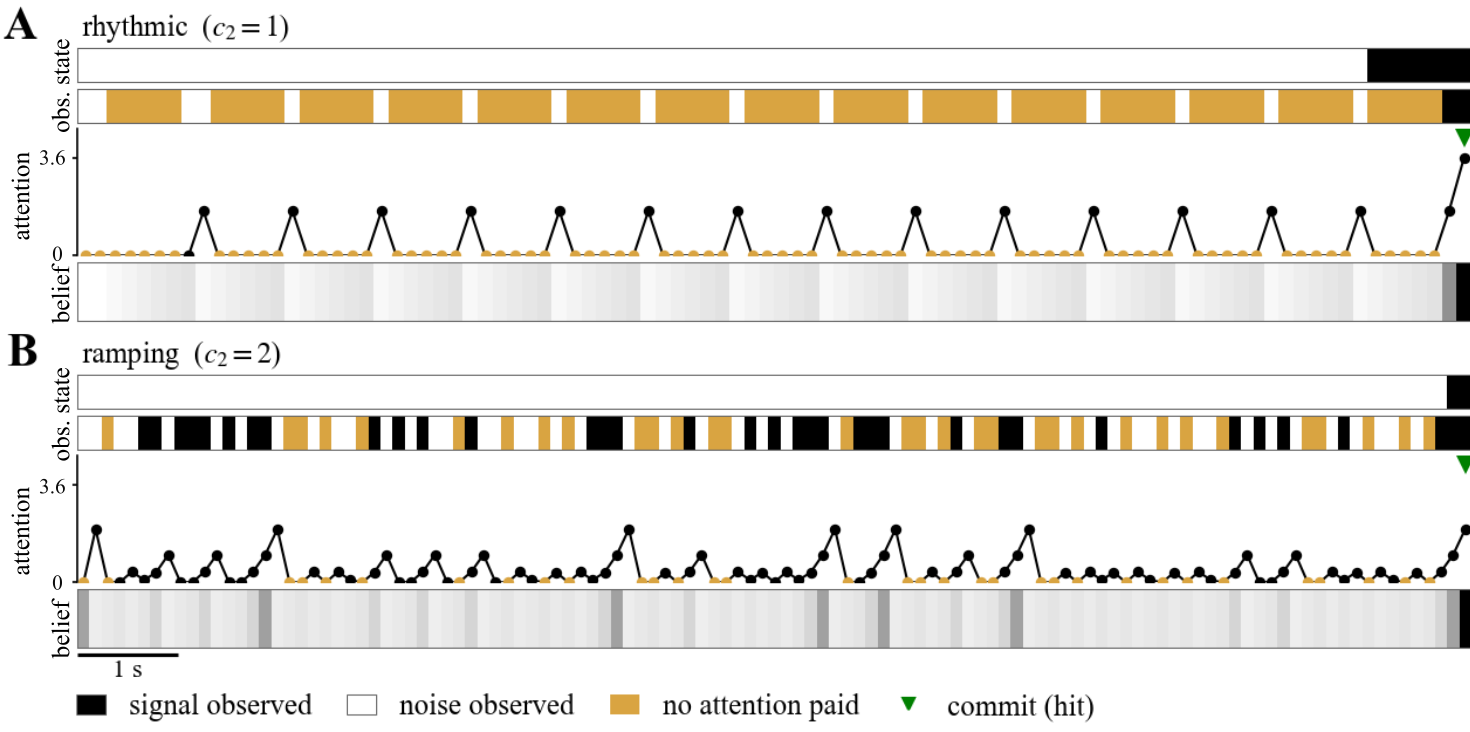}
    \caption{Illustration of agents' interactions with simulated trials. Trial phase (noise: white, signal: black), observations (amber: no attention; white/black: noise/signal observation), belief, and commit action (green triangle). ({\bf A}) agent with a rhythmic policy over a single trial of approximately 100 steps. ({\bf B}) Same as {\bf A} but with a ramping policy.}
    \label{episode}
\end{figure}
 
\textbf{Rhythmic policies} (Fig. \ref{episode}A)\textbf{.} At the start of a trial, when signal belief is 0, the agent pays no attention and waits. Because an observation with 50\% accuracy carries no information, belief during disengagement evolves deterministically under the state transition prior alone. Driven by this prior, the agent's signal belief increases over time until the signal is probable enough to be worth checking. In rhythmic policies, this check is a large installment of attention. The agent is confident in the accuracy of the resulting observation, so if it observes noise, its signal belief drops to near zero, resetting the cycle of wait-then-check. After the first check, the belief just before each check and the belief just after a noise observation are the same every cycle, so the interval between checks is identical every cycle: during the noise period, a rhythmic policy behaves as a fixed-stride schedule, with a stride determined by the task parameters (Section~\ref{prop_sect}). The cycle repeats until a check yields a signal observation, which raises signal belief substantially. Rhythmic policies often have only one jump because, with highly accurate observations, a single signal observation suffices to confirm the signal's presence with high confidence.
 
 
\textbf{Ramping policies} (Fig. \ref{episode}B)\textbf{.} A ramping policy begins the same way, but its first check is a low-attention one. The resulting observation is less accurate and moves belief less, so a signal observation does not justify committing. It does, however, make the signal more probable, which justifies more attention in the next step. Each further signal observation raises attention again, until the agent reaches its highest attention level and uses the observation to decide, or until it observes noise. Because low-attention observations are unreliable, a noise observation lowers belief without resetting it to 0. Without a common reset point, and with a stochastic observation shaping belief at every stage, the regularity of rhythmic policies gives way to irregular ramps. Like a cascade, the agent escalates only while the evidence continues to favor the signal.

Ramping policies have many jumps, one per increment of attention. Some two-jump policies are nonetheless rhythmic: when the first attention level already yields very accurate observations, the agent rarely reaches the second unless the signal is present, so its behavior during noise remains rhythmic.
 
\subsection{The rhythmicity of a policy depends on its attention cost function.}

\begin{figure}
    \centering
    \includegraphics[width=\linewidth]{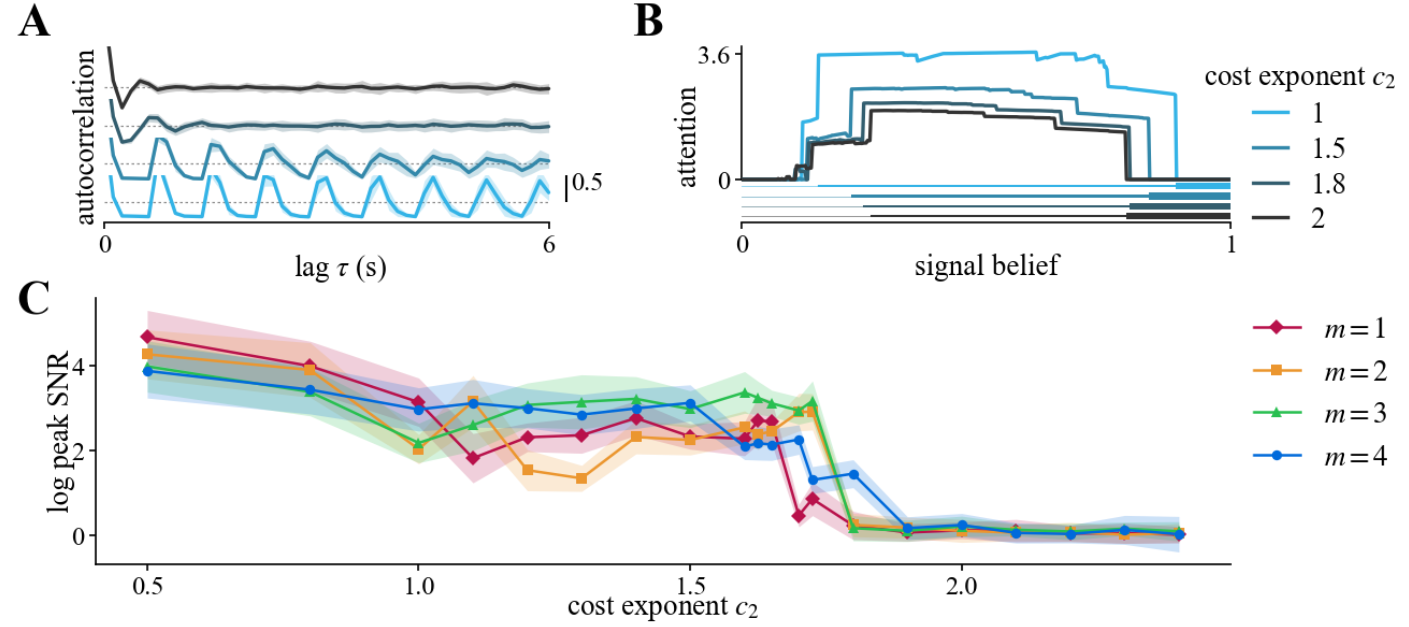}
    \caption{Rhythmicity and policy over attention cost exponent $c_2$. (\textbf{A}) Autocorrelation of attention allocations within a session for optimal policies under the attention cost function $C(a)=0.2a^{c_2}$, for different $c_2$. Rhythmic policies show periodic spikes in autocorrelation; ramping policies are flat. (\textbf{B}) The corresponding policies producing the autocorrelations in \textbf{A}. The lines below show commit behavior (thin: never; medium: when observing signal; thick: always). (\textbf{C}) The rhythmicity metric plotted over $c_2$ for 1, 2, 3, and 4 signal states. Rhythmicity drops sharply at approximately $c_2=1.7$: below this value, policies are rhythmic, and above it, policies ramp.}
    \label{fig:figc2}
\end{figure}
 
The exponent $c_2$ of the attention cost function $C(a)=c_1a^{c_2}$ determines whether the optimal policy is rhythmic or ramping. When the attention cost is linear ($c_2=1$), optimal attention is rhythmic; when it is quadratic ($c_2=2$), optimal attention ramps. The change is visible both in the autocorrelation of attention within trials (Fig. \ref{fig:figc2}A) and in trial visualizations (Fig. \ref{episode}), and we quantify it with a rhythmicity metric (Appendix~\ref{rmetric}) to characterize the relationship between $c_2$ and rhythmicity (Fig. \ref{fig:figc2}C). The transition is sharp: rhythmicity drops abruptly at approximately $c_2=1.7$.
 
When costs are linear or sublinear ($c_2\le1$), a large installment of attention costs no more per unit than a small one, so the agent saves by concentrating attention into sparse, decisive checks, which produces rhythmic policies. When costs are convex ($c_2>1$), the agent saves by spreading attention into smaller installments, but small installments yield less accurate observations. For mildly convex costs, the accuracy of a single large check remains worth its premium, and policies stay rhythmic. Beyond $c_2\approx1.7$, the premium becomes too steep, and ramping policies resolve the cost-accuracy trade-off differently: they keep attention low most of the time, raise it only at critical moments when the signal may be present, and raise it in small increments. By the time an agent with a ramping policy commits, it has observed signal on multiple consecutive occasions, which together constitute stronger evidence than any single low-attention observation.
 
\subsection{Structured attention outperforms constant attention.}

\begin{figure}
    \centering
    \includegraphics[width=\linewidth]{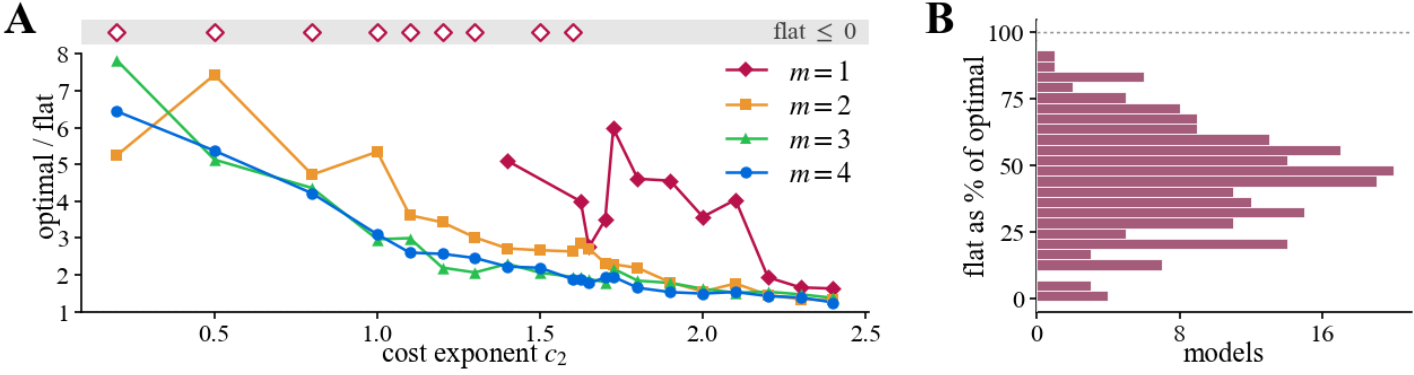}
    \caption{Comparison of optimal vs. constant-attention policies. (\textbf{A}) The ratio, over $c_2$, of mean net reward per step for optimal policies to that of a constant-attention control with the same mean attention. To keep the rest of the plot to scale, we show points where the optimal policy is ten times better, or where the constant policy has a negative net reward, on a grey bar above. (\textbf{B}) The inverse comparison: a histogram, over all solved policies, of how much of an optimal policy's net reward a constant policy with the same mean attention captures. For visual clarity, we show only comparisons where the constant policy achieved non-negative mean reward.}
    \label{fig:fig_gain}
\end{figure}
 
Fluctuating attention is not merely a feature of optimal policies but a substantial source of their value. We compared each optimal policy with a control that allocates constant attention equal to the optimal policy's mean attention.\footnote{For $c_2\neq1$, matching mean attention does not match mean cost. Appendix~\ref{altcontrol} compares optimal policies with a control matched on mean attention cost instead, with comparable results.}
The size of the gain varies with $c_2$ and the other task parameters, but across all parameters tested, 12.45\% of constant-attention controls had negative net reward: allocating the same average attention uniformly rather than selectively yielded a net loss. Among controls with non-negative reward, even the lowest-resolution optimal policies, with 3 and 4 signal states at discretizations of 101 belief levels per state, earned a median of 1.95 and 1.99 times the controls' net reward, respectively. 
The fraction of optimal net reward captured by the controls ranged from 0.7\% to 89.6\%, with a median of 46.4\% and a mean of 45.4\%, and 59.3\% of controls earned less than half the net reward of the corresponding optimal policy (Fig. \ref{fig:fig_gain}). In the language of conditional computation, spending the same budget uniformly forfeits, in the median case, more than half of the achievable net reward.

\subsection{In rhythmic policies, attention frequency depends on other task parameters.}\label{prop_sect}
 
In a rhythmic policy, the stride between checks is itself a function of the task. To characterize this dependence, we regressed the peak frequency of the attention autocorrelation's power spectrum on all swept model parameters jointly by ordinary least squares ($n=181$). We restricted the analysis to models with a rhythmicity metric above 2.0, since the peak frequency of an aperiodic policy reflects only noise, and we report standardized coefficients so that parameters with different swept ranges are comparable.

\begin{figure}
    \centering
    \includegraphics[width=\linewidth]{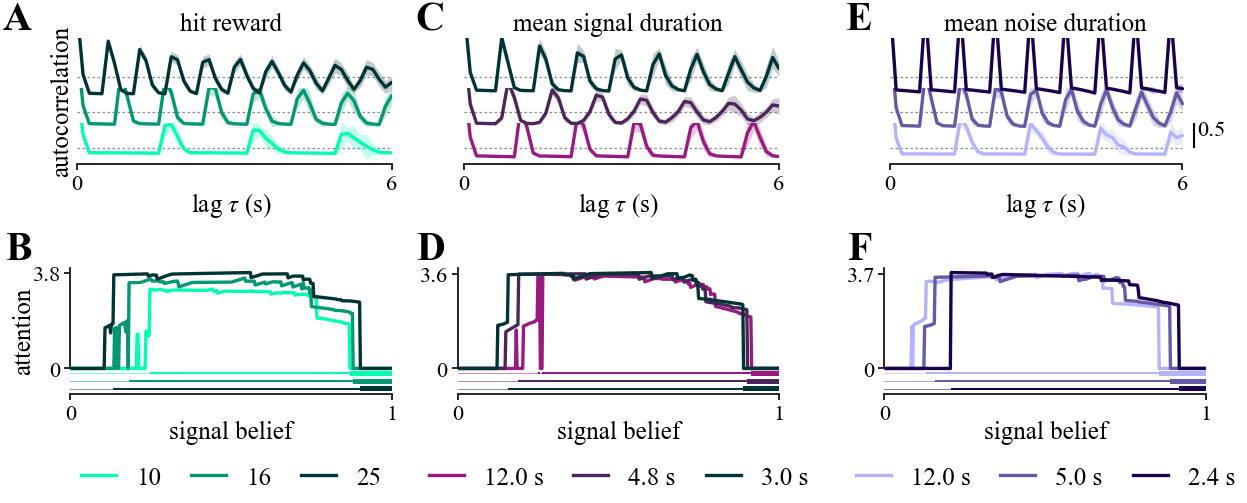}
    \caption{Autocorrelations and policies over model parameters. All models shown here used a single signal state, allowing us to fully visualize their policies. Autocorrelations for models with more signal states are included in Appendix \ref{moresweep}. (\textbf{A}) Autocorrelation of attention allocations within a session for optimal policies under different hit reward values. (\textbf{B}) The corresponding policies producing the autocorrelations in \textbf{A}. (\textbf{C}) and (\textbf{D}): the same for mean signal duration. (\textbf{E}) and (\textbf{F}): the same for mean noise duration.}
    \label{fig:sweep}
\end{figure}
 
\textbf{Effect of signal duration.}
The shorter the mean signal duration, the more frequently the agent attends ($\beta = -.234$, $p < .001$), because it cannot afford to miss a brief signal window (Fig. \ref{fig:sweep}C). The effect is subtle with a single signal state because signal durations are then broadly distributed: a long mean duration does not rule out a brief signal. The effect becomes more pronounced as the number of signal states increases, since this narrows the distribution of signal durations and makes short signals less likely (Appendix~\ref{moresweep}).
 
\textbf{Effect of noise duration.} Similarly, the shorter the mean noise duration, the more frequently the agent attends ($\beta = -.211$, $p < .001$; Fig. \ref{fig:sweep}E). A short mean noise duration means a high probability of transitioning to the signal at each step, so signal belief climbs quickly to the level at which attention pays off. A long mean noise duration means a slow climb and, accordingly, less frequent checks.
 
\textbf{Effect of hit reward.}
Higher hit rewards increase both the frequency of an agent's attention allocations ($\beta = .475$, $p < .001$) and the average attention per allocation (Fig. \ref{fig:sweep}A-B). Greater potential returns justify the cost of more frequent high-quality observations.

\section{Discussion}
 
\textbf{When is a fixed stride optimal?} Our results identify two conditions under which fixed-stride schedules, a common heuristic for allocating computation over streams, are optimal: the cost of attention must be close to linear, and between checks the agent's belief must be driven by its prior rather than by informative observations. When both hold, each decisive check that finds no signal leaves the same belief, so the interval never varies. When either fails, a fixed stride is no longer optimal: steep costs favor graded escalation, and informative observations make the timing of attention depend on what the agent sees. Even when a fixed stride is optimal, the stride should be set by the task rather than treated as a generic hyperparameter: checks should be more frequent when detection is more valuable and when events are brief or frequent (Section~\ref{prop_sect}).
 
\textbf{Cost curvature and cascades.} Ramping policies share their structure with cascades, which escalate from cheap to expensive computation only while evidence remains favorable \citep{viola2001rapid, chen2023frugalgpt}. Our model suggests that the choice between a fixed stride and a cascade-like strategy should depend on how steeply the cost of processing rises with the reliability it buys. This is a property of the whole system, not of compute alone. If reliability shows diminishing returns in compute, as it often does, then each additional increment of reliability costs more than the last, and the effective cost function is convex even when compute is billed linearly. In such settings, our results favor graded escalation over periodic, all-or-nothing checks. Testing this prediction in a deployed system would require measuring its cost-reliability curve, which we leave to future work.
 
\textbf{Temporal structure in correlates of attention.} Our theory predicts how neural and physiological correlates of attention could change with task parameters, including rhythmic allocations or attentional ramps. Several studies have found rhythmic structure in biological attention. In vision, attention is deployed periodically even to a single target \citep{vanrullen2007blinking}, facilitates perception rhythmically \citep{busch2010spontaneous}, is intrinsically rhythmic in its top-down form \citep{helfrich2018neural}, and cycles faster under greater attentional demand \citep{merholz2022periodic}. In audition, performance peaks at a specific beat frequency, suggesting a natural sampling rate for periodic attention \citep{zalta2020natural}, and temporal attention can be allocated selectively for short-term consolidation \citep{shen2011temporal}.

\textbf{Limitations and future work.} Observations in our model are binary, and attention can change instantaneously at no cost; richer observations or switching costs \citep{chebolu2022vigilance} may yield different dynamics. Fatigue could be added as a gradual increase in $c_1$. Value iteration makes our results exact up to discretization, but the belief space grows with $m$, limiting us to $m\le4$. In preliminary checks, point-based POMDP solvers and an independent value iteration implementation recovered both regimes, whereas deep reinforcement learning agents did not do so reliably. We suspect this reflects the approximation error of learned policies rather than the regimes themselves, but whether learned policies can recover them remains an open question. Finally, comparing optimal policies with other heuristics, such as the best fixed-stride schedule in the ramping regime, would quantify the cost of choosing the wrong strategy.
 
\textbf{Broader impacts.} This work contributes to the theoretical understanding of how biological and artificial agents should allocate costly attention. In neuroscience, it offers a normative framework for interpreting fluctuations in attention. In machine learning, it may inform the design of systems that must process large or continuous inputs under computational budgets. These systems face an equivalent trade-off: applying maximum processing everywhere is prohibitively expensive, but sparse processing risks missing relevant content. Our framework predicts that the optimal strategy in such settings depends on the cost structure of the processing pipeline.
 
\section{Conclusion}
We proposed a normative model of sustained attention with explicit attention costs and solved it by value iteration. Optimal agents either attend rhythmically, in sparse, decisive checks at regular intervals, or ramp attention as evidence accumulates, depending chiefly on how steeply attention costs grow. These results offer a normative account of attention fluctuations in sustained attention tasks and a basis for deciding when a system that allocates computation over a stream should use a fixed stride and when it should escalate adaptively.

\section*{AI, Ethics, and Reproducibility Statements}

\textbf{AI use statement.} In this work, we used generative AI tools to revise the manuscript, to identify related literature in machine learning, and to improve code. All AI-modified text was reviewed and edited by the authors, and all suggested references were checked by the authors against the original sources. In the code, we used AI tools to optimize the value iteration algorithm for shorter run-times and to augment the code we wrote to generate figures---re-organizing the code for more modularity and modifying the code to reformat figures. 
 
We did not use generative AI tools to formulate our model, to analyze our results, to prove or assist in proving any mathematical claims, or to propose hypotheses/research questions.
 
\textbf{Ethics statement.} We do not believe the paper violates any part of the ICLR Code of Ethics.
 
\textbf{Reproducibility statements;} We plan to publicly release code for building, training, simulating, and assessing the models on GitHub.
 
\textbf{Acknowledgments} This work was supported in part by NSF CAREER grant 1552868 to XP, by AFOSR grant FA9550-21-1-0422 in the Cognitive and Computational Neuroscience program to XP, and by NIH grants R03DC015618 and R01DC017797 to MM.
 
\bibliographystyle{plainnat}
\bibliography{Styles/draft_ML.bib}
 
 
\appendix
 
\section{Appendix}\label{appendix}
 
\subsection{Signal Detection Model Defaults}
 
The default probability $q$ of transitioning from the noise state to the first signal state was 2.4\%, for a mean noise duration of approximately 42 time steps. The default mean signal duration was 25 time steps, so each signal state transitions out with probability $p=\frac{m}{25}$. For the default attention cost function, we used $c_1=0.2$ and $c_2=1.0$. For the default outcome rewards, we used a hit reward of 20, a commit cost of 5, and no explicit miss or false alarm penalty. As a result, the effective hit reward is 15, the difference between the reward collected for a correct commit and the commit cost, while the false alarm penalty is 5 and a miss leads to neither a reward nor a penalty. Our default temporal discount factor was 0.995, and we used an inter-trial interval of 16.67 to 25 time steps, which encouraged agents to perform well within the current trial without neglecting future trials. Rhythmic and ramping policies are optimal both with and without this inter-trial interval. 
 
We took inspiration for these task parameters from \citet{de2022strategic}, in which mice performed a sustained attention task for many consecutive trials in each session. With time steps of 120 milliseconds, many of our default parameters match the timing of their experimental task, in which noise durations were exponentially distributed with a mean of 5 seconds, signals lasted 3 seconds, and inter-trial intervals ranged from 2 to 3 seconds. 
We used different discretizations for different values of $m$. For $m=1$, we ran value iteration on 10001 evenly spaced signal beliefs. For $m=2$, we used every combination of 1001 probability levels in the first signal state and 1001 in the second, excluding pairs whose sum exceeded 1. For $m=3$ and $m=4$, we used 101 probability levels per signal state.

\subsection{Signal Detection Model Solution Through Value Iteration}\label{vimethods}
 
Because the belief $b$ is a sufficient statistic for the agent's history, we solve the POMDP as a fully observed MDP over beliefs. Within a step, the agent chooses an attention level $a\ge0$ and a commit rule $\kappa:\{\text{noise},\text{signal}\}\to\{0,1\}$ specifying whether to commit after each possible observation. The optimal value function satisfies
\begin{equation}
V(b)=\max_{a\ge0,\;\kappa}\Big\{-C(a)+\sum_{o}P(o\mid b,a)\Big[\kappa(o)\,R(b^{o})+\big(1-\kappa(o)\big)\,\gamma\,V\big(\tau(b^{o})\big)\Big]\Big\},
\label{eq:bellman}
\end{equation}
where $b^{o}$ is the posterior belief after observing $o$, $\tau$ propagates a belief through the transition dynamics, and $R(b^{o})$ is the expected reward for committing under belief $b^{o}$: the hit reward weighted by the posterior probability of the signal, minus the commit cost. Committing additionally yields the discounted value of subsequent trials after the inter-trial interval, which we omit from Eq.~\ref{eq:bellman} for clarity.
 
We solve Eq.~\ref{eq:bellman} by value iteration on the belief grid described above. For any belief at the start of a step, the expected reward within that step is an analytic function of attention, which we maximize to obtain the optimal attention and its value. In each subsequent iteration, the value function from the previous iteration serves as the discounted continuation value, so that after $n$ iterations the solution looks $n$ steps into the future. We iterate either until convergence, when optimal attention is identical between steps, or until $n=919$, when looking further into the future confers almost no additional benefit due to temporal discounting ($0.995^{919}<0.01$). For off-grid beliefs, optimal attention and expected value are computed from the grid via linear interpolation.
 
\subsection{Rhythmicity Metric}\label{rmetric}
 
To analyze the relationship between task parameters and the rhythmicity of the resulting policies, we use a numerical metric for rhythmicity. We compute the within-trial autocorrelation of attention from trials simulated under the optimal policy and take the log of the signal-to-noise ratio at the peak of the autocorrelation's power spectral density. A sharp spectral peak yields a higher log ratio and indicates consistent rhythmic structure; a broad, featureless spectrum indicates aperiodic or ramping allocation.
 
\subsection{Other policy visualizations}\label{morevis}
 
\begin{figure}[h]
    \centering
    \includegraphics[width=\linewidth]{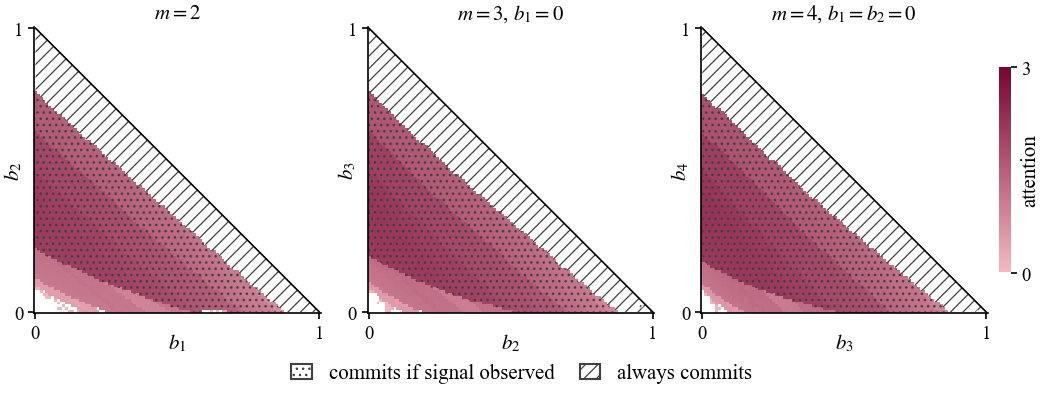}
    \caption{We may also visualize policies over two dimensions, allowing a look at how attention should be allocated over belief distributed in two signal states. For $m=2$, this covers the full policy space. For $m>2$, this enables visualizations of the policy over belief in two of the signal states while the others are fixed at 0. The same ``jumps'' in optimal attention can be observed here in the diagonal bands.}
    \label{fig:A1-1}
\end{figure}

In addition to one-dimensional policy visualizations, which plot optimal attention on the y-axis against signal belief on the x-axis, we can plot two-dimensional visualizations that show optimal attention as color over belief in two signal states, one on each axis (Fig. \ref{fig:A1-1}). Attention jumps are evident in these visualizations as well.
 
\begin{figure}[h]
    \centering
    \includegraphics[width=\linewidth]{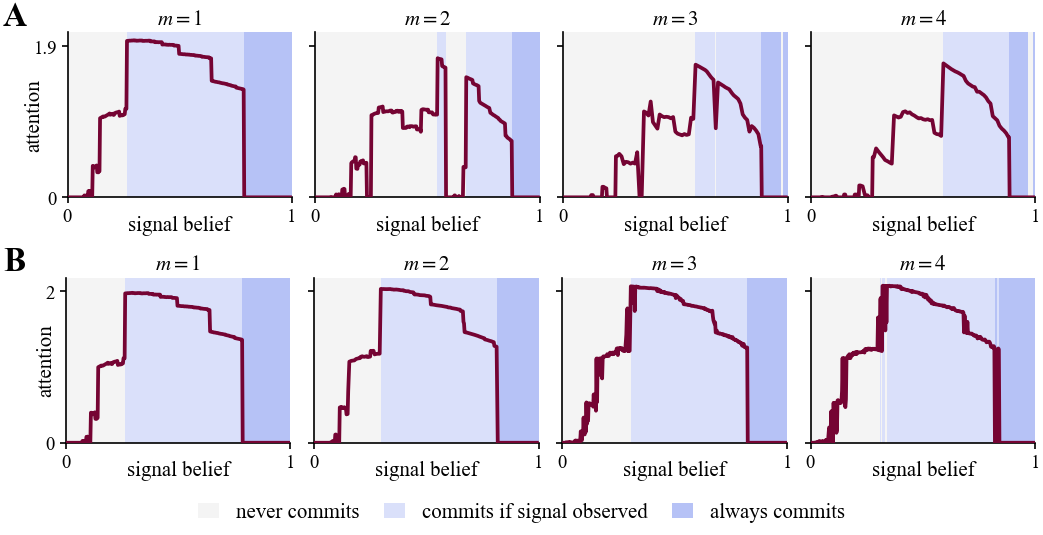}
    \caption{Visualization of policies and policy slices, as in Fig. \ref{fig:policy1}, except that signal belief in the slices is (\textbf{A}) concentrated in the first signal state or (\textbf{B}) distributed uniformly across all modeled signal states.}
    \label{fig:A1-2}
\end{figure}

In the main text, we focus on policy slices---in which we vary only total signal belief to show a slice of a policy over a higher-dimensional belief space---where total signal belief is concentrated in the last signal state. We made this choice because it yields the cleanest visualizations and thus conveys the consistency of attention jumps most efficiently. However, we also visualized policy slices where signal belief is concentrated in the first signal state (Fig. \ref{fig:A1-2}A) and where signal belief is spread uniformly across all signal states (Fig. \ref{fig:A1-2}B). These slices all exhibit jump-like behavior, but the overall structure of the higher-dimensional policy is less cleanly discernible. For instance, one can easily see bands of different attention levels in the $m=2$ two-dimensional visualization (Fig. \ref{fig:A1-1}, left), while the policy slice concentrated on the first signal state, for this exact same policy, appears erratic (Fig. \ref{fig:A1-2}A; $m=2$). 
 
We postulate that policy slices appear more erratic for signal belief concentrated in earlier signal states for the following reason. When signal belief is concentrated in earlier states, there are more options for optimal strategies. The signal does not end until the world transitions out of the last signal state, so if the agent thinks the last signal state is unlikely or impossible, it does not need to act urgently to collect reward. It could wait, attend only slightly to refine its estimate, or attend a lot to decide whether it can collect an early reward. Many of these options may be equally good. When signal belief is concentrated in the last signal state, by contrast, waiting and slight refinements lose value compared to commitment decisions, and the balance of good vs. bad performance becomes more immediate, with fewer opportunities to act anticipated in the near future. This may restrict the possible optimal levels of attention. Optima over beliefs concentrated in early signal states may therefore have ``noise'' introduced, where one of two equally good options is selected for one exact belief and the other is selected for a similar, adjacent belief. Viewing the policy over these adjacent beliefs results in a pattern that looks random. By contrast, there may be few optima, or only one, over beliefs concentrated in the late signal states, resulting in more consistency between the optimal attention at adjacent beliefs. The central point made in the paper, that policies exhibit jumps, holds whether or not this explanation is correct.
 
\subsection{Autocorrelations for \texorpdfstring{$m\ge1$}{m >= 1}}\label{moresweep}
 
\begin{figure}[htp]
    \centering
    \includegraphics[width=\linewidth]{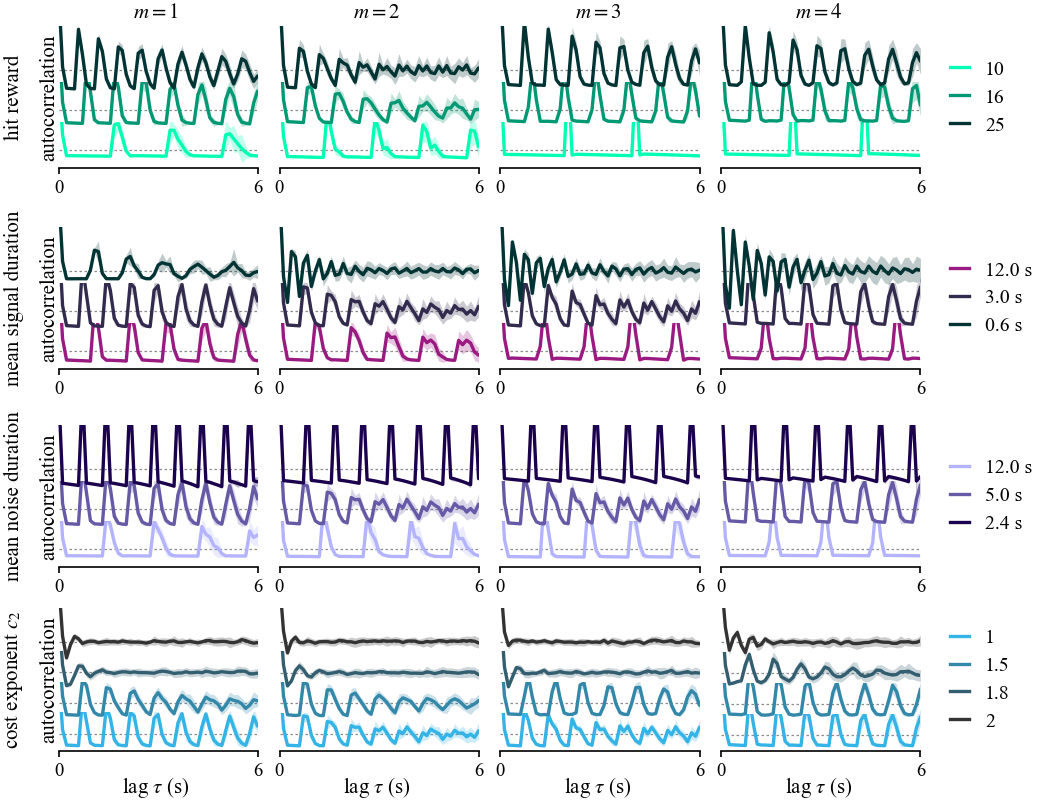}
    \caption{Autocorrelations over model parameters. The columns show models with $m=1$, 2, 3, and 4, respectively. The rows show autocorrelations over different values of the following parameters, from top to bottom: hit reward, mean signal duration, mean noise duration, temporal discount factor, and attention cost exponent.}
    \label{fig:sweep2}
\end{figure}
Fig. \ref{fig:sweep2} shows autocorrelations for policies representing the signal with different numbers of signal states. Results hold as $m$ increases, and the effects of varying temporal parameters generally become easier to read for higher $m$. As $m$ increases, allocation frequency increases for short signal durations and decreases for long signal durations, consistent with the expected effect of sharpening the duration distribution. Across all $m$, increases in hit reward similarly increase allocation frequency, and linear attention costs yield rhythmicity, while quadratic costs do not.

\subsection{Flat controls with equal mean attention cost instead of equal mean attention}\label{altcontrol}

\begin{figure}
    \centering
    \includegraphics[width=\linewidth]{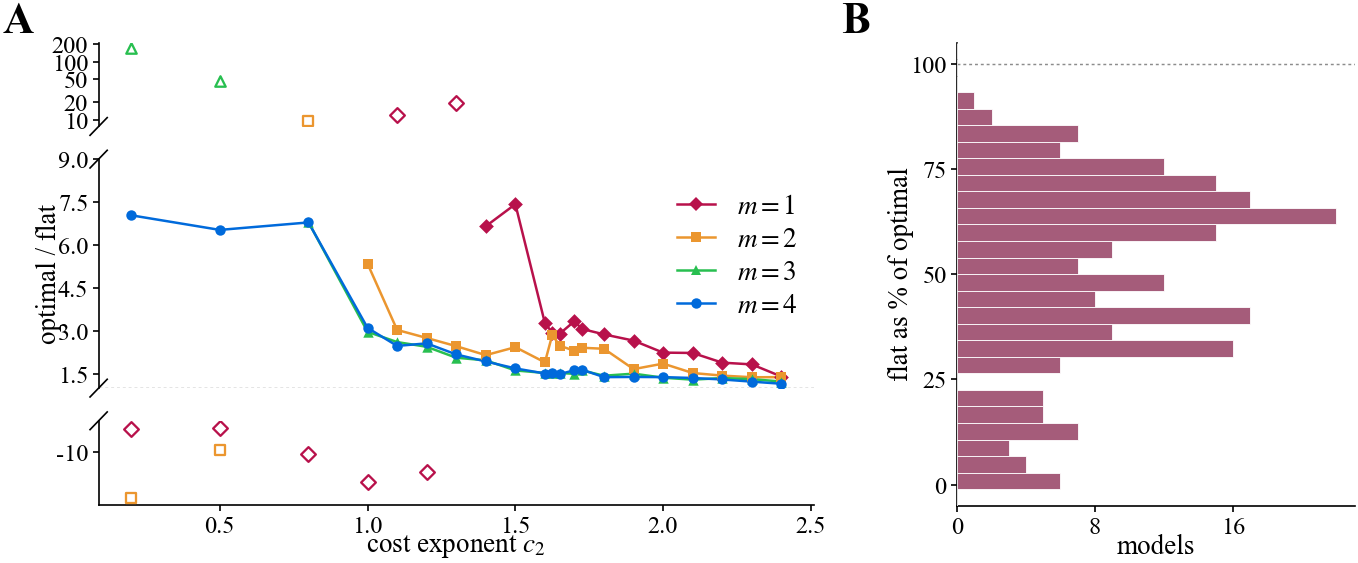}
    \caption{Comparison of optimal vs.\ cost-matched constant-attention policies. (\textbf{A}) The ratio, over $c_2$, of mean net reward per step for optimal policies to that of a constant-attention control with the same mean attention cost. A broken axis shows negative ratios and ratios above 10. (\textbf{B}) The inverse comparison: a histogram, over all solved policies, of the percentage of an optimal policy's net reward captured by a constant-attention control with the same mean attention cost. For visual clarity, we show only comparisons where the control achieved non-negative mean reward.}
    \label{fig:gain2a}
\end{figure}

Because attention cost depends nonlinearly on attention whenever $c_2\neq1$, matching a control's mean attention does not match its mean cost. We therefore also compared optimal policies with a second constant-attention control, whose attention level was chosen so that its mean attention cost matched that of the optimal policy. Results were comparable to those for the mean-attention control (Fig. \ref{fig:gain2a}). Across all parameters tested, 11.62\% of cost-matched controls had negative net reward. Among controls with non-negative reward, the lowest-resolution optimal policies ($m=3$ and $m=4$) earned a median of 1.54 and 1.61 times the controls' net reward, respectively. The fraction of optimal net reward captured by the controls ranged from 0.6\% to 89.6\%, with a median of 54.7\% and a mean of 50.2\%, and 46.4\% of controls earned less than half the net reward of the corresponding optimal policy.

The largest advantages of optimal policies were greater against the cost-matched control than against the mean-attention control. For the base parameter set, the maximum ratio of optimal to control net reward against the mean-attention control was 5.97, 7.43, 7.83, and 7.80 for $m=1$, 2, 3, and 4, respectively, compared with 19.77, 9.64, 171.69, and 8.47 against the cost-matched control.

\end{document}